# Signal-to-Noise Eigenmode Analysis of the Two-Year COBE Maps

J. Richard Bond

*Canadian Institute for Theoretical Astrophysics, University of Toronto, Toronto, Ontario, Canada M5S 1A7*

To test a theory of cosmic microwave background fluctuations, it is natural to expand an anisotropy map in an uncorrelated basis of linear combinations of pixel amplitudes — statistically-independent for both the noise and the signal. These $S/N$-eigenmodes are indispensible for rapid Bayesian analyses of anisotropy experiments, applied here to the recently-released two-year COBE *dmr* maps and the *firs* map. A 2-parameter model with an overall band-power and a spectral tilt $\nu_{\Delta T}$ describes well inflation-based theories. The band-powers for *all* the *dmr* 53, 90, 31 $a+b$ GHz and *firs* 170 GHz maps agree, $\{(1.1 \pm 0.1) \times 10^{-5}\}^{1/2}$, and are largely independent of tilt and degree of (sharp) $S/N$-filtering. Further, after optimal $S/N$-filtering, the *dmr* maps reveal the same tilt-independent large scale features and correlation function. The unfiltered *dmr* 53 $a+b$ index $\nu_{\Delta T} + 1$ is $1.4 \pm 0.4$; increasing the $S/N$-filtering gives a broad region at $(1.0$–$1.2)\pm0.5$, a jump to $(1.4$–$1.6)\pm0.5$, then a drop to 0.8, the higher values clearly seen to be driven by $S/N$-power spectrum data points that do not fit single-tilt models. These indices are nicely compatible with inflation values ($\sim 0.8$–$1.2$), but not overwhelmingly so.

PACS NOs: 98.80.Cq, 98.80.Es, 98.70.Vc

The importance of the large-angle COBE *dmr* [1,2] and the balloon-borne *firs* [3] detections for testing theories of cosmic structure formation can hardly be overstated. For theories in which the multipole components of the radiation pattern $a_{\ell m}$ are Gaussian-distributed and statistically-independent, as in inflation-based models or *via* the central limit theorem, we only need to determine the power spectrum, $\mathcal{C}_\ell \equiv \ell(\ell+1)\langle|a_{\ell m}|^2\rangle/(2\pi)$. A phenomenology characterized by just two parameters, a broad-band power $\langle \mathcal{C}_\ell \rangle_{dmr}$ and a broad-band tilt $\nu_{\Delta T}$ is an excellent approximation for a large class of models:

$$\mathcal{C}_\ell = \langle \mathcal{C}_\ell \rangle_{\overline{W}} \frac{\mathcal{U}_\ell\, \mathcal{I}[\overline{W}_\ell]}{\mathcal{I}[\overline{W}_\ell \mathcal{U}_\ell]}\,,\ \mathcal{U}_\ell \equiv \frac{\Gamma\left(\ell+\frac{\nu_{\Delta T}}{2}\right)\Gamma(\ell+2)}{\Gamma(\ell)\Gamma\left(\ell+2-\frac{\nu_{\Delta T}}{2}\right)}\,,$$

$$\langle \mathcal{C}_\ell \rangle_{\overline{W}} \equiv \frac{\mathcal{I}[\mathcal{C}_\ell \overline{W}_\ell]}{\mathcal{I}[\overline{W}_\ell]}\,,\ \text{where}\ \mathcal{I}[f_\ell] \equiv \sum_\ell \frac{(\ell+\tfrac{1}{2})}{\ell(\ell+1)} f_\ell$$

defines the "logarithmic integral" of a function $f_\ell$. For the *dmr* and *firs* maps, the broad-band filter $\overline{W}_\ell$ encodes beam-smearing and effects from pixelization [9]; for other experiments it also encodes switching information. The virtue of the band-power [4–6] $\langle \mathcal{C}_\ell \rangle_{\overline{W}}$ is that it is insensitive to the specific shape of $\mathcal{C}_\ell$, whereas the oft-used [1] "$Q_{rms,PS}$" varies considerably:

$$\frac{Q_{rms,PS}}{17.6\mu K} \approx \frac{\langle \mathcal{C}_\ell \rangle_{\overline{W}}^{1/2}}{10^{-5}} e^{-\alpha \nu_{\Delta T}(1+0.3\nu_{\Delta T})}\,,\ \alpha = 0.37\ dmr;$$

$\alpha = 0.55$ for *firs*. The broad-band tilt $\nu_{\Delta T}$ measures deviation from scale-invariance in $\Delta T$: $\mathcal{U}_\ell = (\ell+\tfrac{1}{2})^{\nu_{\Delta T}}(1+\mathcal{O}((\ell+\tfrac{1}{2})^{-2})$. It is usual to use $n_s = \nu_{\Delta T} + 1$ to characterize the slope [1], because for pure scalar metric fluctuations in a cold dark matter Universe with almost no baryons, $n_s$ is the primordial (*e.g.*, post-inflation) index [7]. However, the standard 5% baryon content gives $\nu_{\Delta T} = 0.15$ for an initially untilted primordial spectrum [8,5]; and primordial tilts, usually negative, can arise from gravitational wave as well as density fluctuation modes, but with $n_s > 0.7$ required to get the observed cluster abundance (*e.g.*, [8,6]).

The observations $(\Delta T/T)_p$ from the $p^{th}$ pixel of an $N_{pix}$ CMB anisotropy experiment is given to us in terms of an average value $\overline{(\Delta T/T)_p}$ and a correlation matrix $C_{Dpp'}$, with possibly off-diagonal components as well as the diagonal $\sigma_{Dp}^2$. The theory we are testing is characterized by a correlation matrix $C_{Tpp'}$ (here $= \mathcal{I}[\mathcal{C}_\ell \overline{W}_\ell P_\ell(\cos\theta_{pp'})]$, where $\theta_{pp'}$ is the angle between pixels $p$ and $p'$). If signals are not Gaussian-distributed, higher order correlation tensors are required, and such theories can only be well-analyzed by simulation.

For this exploration of the *dmr* maps, I used: one lower resolution scale than the original maps, $5.2°$ *c.f.* $2.6°$; a Galactic latitude cut $|b| > 25°$ (leaving 928 pixels); dipole and average subtractions were done after cuts, but before resolution-lowering; the (non-Gaussian) revision of the *dmr* beam [9], with corrections for digitization and pixelization; a correction linear in the off-diagonal $C_D$ components to test sensitivity to residual correlation because COBE actually measures a $\Delta T$ between pixels $60°$ apart. For the 168 GHz *firs* map, $2.6°$ rather than $1.3°$ pixels were used (but I checked that the two give the same answer), leaving 1070 pixels; a $3.9°$ beam was used, which includes pixelization corrections.

A full Bayesian analysis of maps requires frequent inversion and determinant evaluations of $N_{pix} \times N_{pix}$ correlation matrices, the sum of all $C_{Tpp'}$ in the theoretical modelling plus the pixel-pixel observational error matrix $C_{Dpp'}$. To facilitate this, I expand the pixel values $(\Delta T/T)_p$ into a basis of "signal-to-noise" eigenmodes for the maps in which the transformed noise and transformed (wanted) theoretical signal we are testing for do not have mode-mode correlations, *i.e.*, are orthogonal. This can always be done, no matter what the experiment [4]. Complications are associated with the removal of averages, dipoles, *etc.* and the existence of secondary signals in the data, both of which do couple the modes. A model for the various contributions that make up the observed data is then

$$\xi_k = \sum_{p=1}^{N_{pix}} (RC_D^{-1/2})_{kp} (\Delta T/T)_p$$
$$= s_k + (1+r)n_k + c_k + \text{res}_k, \ k=1,\ldots,N_{pix},$$
$$\langle s_k s_{k'}\rangle = \mathcal{E}_{TR,k}\delta_{kk'} = \left(RC_D^{-1/2} C_T C_D^{-1/2} R^\dagger\right)_{kk'},$$
$$\langle n_k n_{k'}\rangle = \delta_{kk'}, \ \langle \text{res}_k \text{res}_{k'}\rangle = \mathcal{R}_{kk'}, \ \langle c_k c_{k'}\rangle = \mathcal{K}_{kk'},$$
$$(\mathcal{E}_{totR})_{kk'} \equiv \mathcal{E}_{TR,k} + (1+r)^2 \delta_{kk'} + \mathcal{R}_{kk'} + \mathcal{K}_{kk'},$$

where $R$ is a rotation matrix. The modes are sorted in order of decreasing $S/N$-eigenvalues, $\mathcal{E}_{TR,k}$, so low $k$-modes probe the theory in question best. With uniform weighting and all-sky coverage, the $S/N$-modes are just the independent $Re(a_{\ell m})$ and $Im(a_{\ell m})$, with the lowest $\ell$ having the highest $\mathcal{E}_{TR,k}$, hence $k \sim (\ell+1)^2$. With inhomogeneous pixel coverage, Galactic cuts, dipole subtractions, *etc.*, they are more complex.

This expansion is a complete (unfiltered) representation of the map. The sum of $\xi_k^2$ over bands in $S/N$-space defines a $S/N$-power spectrum which gives a valuable picture of the data; an example is shown in Fig.1. The maps can have an arbitrary average, dipole and possibly quadrupole (because the Galactic contribution may contaminate the signal's quadrupole). This is modelled by $c_k$, with 4 or 9 components, which I take to be Gaussian-distributed with very wide width, *i.e.*, with a uniform prior probability. In $S/N$-space, the variance $\mathcal{K}_{kk'}$ has off-diagonal terms affecting the largest scale $S/N$-modes and complicating the interpretation of Fig.1 for small $k$, but in the Bayesian analysis these unknowns can immediately be marginalized (by integrating over all $c_k$).

Fig.1 also shows anomalies relative to the power-law theory at high $k$ (see [5] for more examples). I model these residuals by an excess pixel noise with an amplification factor $r$. For *firs*, $r \sim 0.25$. For the 2-year data, the most likely $r < 1\%$ for the recommended *dmr* noise level for all maps, often zero for $\nu_{\Delta T} > 0.2$. For *dmr*, I include it anyway because of the high sensitivity of the full Bayesian method, but the results differ very little if I set $r = 0$. Features in Fig.1 that do not fit the smooth theoretical model (around $k \sim 200$ for 53a+b) represent unknown components, denoted by $\text{res}_k$, which could point to a better model power spectrum to try, or residuals left over from the various *dmr* map-processing techniques. In general, the $\text{res}_k$ covariance $\mathcal{R}$ would be nondiagonal. The nature of the low and high $S/N$-modes has been probed with angular correlation functions [6]. These reveal the expected result that high $S/N$-modes do indeed involve collective large scale pixel combinations, while low $S/N$-modes involve destructive interferences from nearby pixels that are insensitive to large scale structure in the maps, but are quite sensitive to physics inside the beam, whether from systematic effects or true 'white' noise on the sky. Thus $S/N$-modes form an ideal set for filtering.

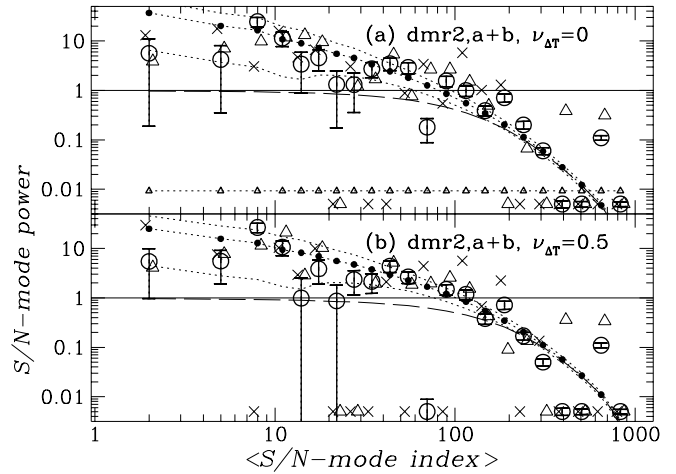

FIG. 1. The $(S/N)$ band-power (mean of $\bar{\xi}_k^2 - 1$ over a band in $k$) for the *dmr* 2-year 53a+b maps, using the (a) $\nu_{\Delta T}=0$ and (b) $\nu_{\Delta T}=0.5$ modes. The observational error bars (smaller than the points for $k > 30$) include a large scale power correction. The $<0.01$ values at high $k$ are actually slightly negative. The small solid points joined by the central dotted curve give band-averages of $\mathcal{E}_{TR,k}$, with the upper and lower dotted curves giving theoretical variances for this binning; all three scale with $\langle \mathcal{C}_\ell \rangle_{dmr}$, thus can be raised or lowered to best fit the data. The horizontal line of solid triangles shows the most likely residual noise (tiny for (a), off-scale for (b)). The Weiner-filter is the long-dashed curve. The 90a+b (open triangle) and 31a+b ($\times$) data points shown are uniformly scaled until their $\mathcal{E}_{TR,k}$ agrees with 53a+b (which they do): thus, given the larger error bars, the agreement in power among maps is reasonably good.

The first step in the Bayesian method is the construction of a joint likelihood function in $\langle \mathcal{C}_\ell \rangle_{dmr}^{1/2}$, $\nu_{\Delta T}$ and $r$ (adopting a uniform prior probability in each of these variables). Integrating over $r$ (marginalizing it) allows one to construct $\nu_{\Delta T}$-$\langle \mathcal{C}_\ell \rangle_{dmr}^{1/2}$ contour maps. Marginalization over $\nu_{\Delta T}$ or cuts along a given $\nu_{\Delta T}$ (*e.g.*, 0) gives a $\langle \mathcal{C}_\ell \rangle_{dmr}^{1/2}$ distribution; marginalization over $\langle \mathcal{C}_\ell \rangle_{dmr}^{1/2}$ gives a $\nu_{\Delta T}$ distribution. The Table gives 50% Bayesian probability values for $\langle \mathcal{C}_\ell \rangle_{dmr}^{1/2}$ and $\nu_{\Delta T}$, with 'one-sigma' error bars from the 84% and 16% Bayesian values, 'two-sigma' from the 97.5% and 2.5% values [11]. The $\langle \mathcal{C}_\ell \rangle_{dmr}^{1/2}$ for all of the 53, 90 and 31 GHz *dmr* maps and *firs* map agree at better than the one-sigma level, along fixed $\nu_{\Delta T}$ lines, among different $\nu_{\Delta T}$, and when $\nu_{\Delta T}$ is marginalized, essentially independently of the degree of $S/N$-filtering (cuts at 400 and 200 are shown). There is also no discernable power in the *dmr* a−b maps. These band-powers compare with: the original first year *dmr* number, us-ing correlation functions (CF) [1], 0.97±.28; a later update [12], 0.97±.16 using the Boughn-Cottingham single quadratic statistic (for $\nu_{\Delta T}=0$); the two-year data, using CF [2], 1.00±.10 (for $\nu_{\Delta T}=0$) and $1.02^{+.43}_{-.27}$ (for the most probable $\nu_{\Delta T}$); and the *firs* CF result [15], 1.08±0.3.

The tilts shown in the Table (with one and two sigma errors) are not determined with the same precision, nor can one attach the same confidence to the values, because of the variation with filtering. In particular, the high point for the 53a+b $S/N$-power in Fig.1 just below $k = 200$ (which comes from 53a and not from 53b) drives the index up somewhat, as do the two slightly high points around 100 (which 90a+b has also), while the low point at 70 drives it down. Thus the most probable index is sensitive to filtering, as Fig.1 makes clear; it also shows a high index power law is *not* a particularly good description of the excess power. Another significant feature is that the low quadrupole drives the index higher (the '+q' results). I do not actually have to subtract the quadrupole since the large uniform prior allows the analysis to attribute what it cannot fit to $c_k$. I agree with the *dmr* team [14] that the lower indices obtained by assuming Galactic interference in the quadrupole are the better numbers to adopt. Decreasing the beam-size also lowers $\nu_{\Delta T}$ a bit [6]. The $\nu_{\Delta T}$ results compare with: the original first year *dmr* value [1], $1.2^{+0.5}_{-0.7}$; the two-year CF result [2] $1.6^{+0.5}_{-0.6}$; power spectrum (PS) values using quadratic estimators [13] for one-year, $1.7^{+0.5}_{-0.5}$, and two-year, $1.5^{+0.4}_{-0.4}$ (for $3 \le \ell \le 19$), $1.3^{+0.4}_{-0.4}$ (for $3 \le \ell \le 30$), and using Bayesian methods with linear multipole filtering [14], $1.1^{+0.3}_{-0.3}$, (for $3 \le \ell \le 30$). For the first year 53a+b map, I get $1.8^{+0.5;0.8}_{-0.5;1.1}$ with no quadrupole, $2.0^{+0.4;0.7}_{-0.4;1.0}$ with, because the $k \sim 200$ structure is more pronounced [6]. The *firs* CF value [15] is $1.0^{+0.4}_{-0.5}$ including the quadrupole, whereas I get $1.8^{+0.6;1.0}_{-0.8;1.5}$ for *firs*,+q, and $res_k$ is large enough that direct $S/N$-filtering can give even higher values [6].

I have found sharp $S/N$-filtering preferable for statistical analysis [16], but smooth preferable for cleaning noise to look for robust map-independent features. Fig.2 shows optimally-filtered maps and their temperature autocorrelation functions, $C(\theta)$. The maps are equal area projections [17] about the pole. Floating averages, dipoles and quadrupoles can be added to bring them into better agreement. This Weiner-filtering is an immediate byproduct of the $S/N$-eigenmode expansion [5,18]: given observations $\bar\xi_k$, the mean value and variance matrix of the desired signal $s_k$ are [19]

$$\langle s_k | \bar\xi \rangle = \sum_{k'} \left\{ \mathcal{E}_{TR} \mathcal{E}_{totR}^{-1} \right\}_{kk'} \bar\xi_{k'},$$

$$\langle \Delta s_k \Delta s_{k'} | \bar\xi \rangle = \left\{ \mathcal{E}_{TR} \mathcal{E}_{totR}^{-1} (\mathcal{E}_{totR} - \mathcal{E}_{TR}) \right\}_{kk'}.$$

The mean field $\langle s_k | \bar\xi \rangle$ is the maximum entropy solution. The operator multiplying $\bar\xi_k$ is the Weiner filter. The fluctuation, $\Delta s_k = s_k - \langle s_k | \bar{\xi} \rangle$, of the signal about the mean is realized by multiplying a vector of $N_{pix}$ independent Gaussian random numbers by the square root of the variance matrix.

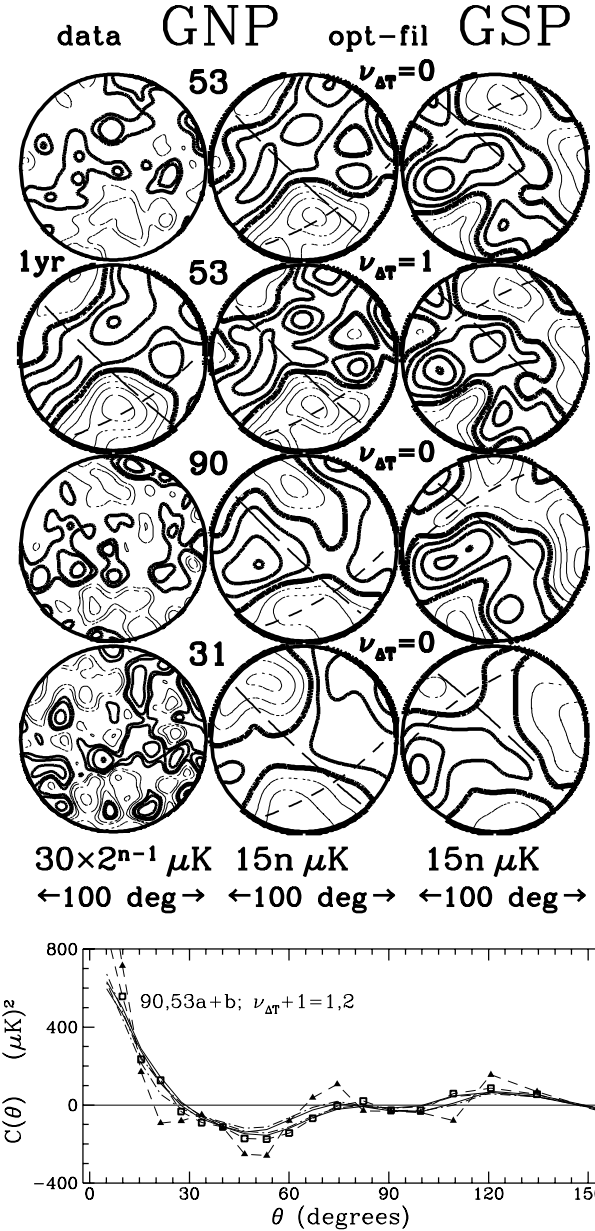

FIG. 2. The first column shows unfiltered 100° diameter *dmr* maps centered on the North Galactic Pole, the second column shows them after Weiner-filtering, the third the South Pole version, assuming a $\nu_{\Delta T}=0$ $\mathcal{C}_\ell$-spectrum, with the $n$th contour as noted, positive contours heavier than negative ones and the zero contour heaviest of all. The $x$-axis points to the Galactic center, the $y$-axis to $\hat{q}_{GNP} \times \hat{q}_{GC}$, the Supergalactic plane is long-dashed, and the ecliptic plane is short-dashed; all are uncorrelated with the features. The second row entries are: the first-year Weiner-filtered GNP result; two-year GNP, GSP results, but with $\nu_{\Delta T} = 1$. The maps have been smoothed by a 3° Gaussian filter. In the lower figure, correlation functions for 53a+b GHz (open circles) and 90a+b GHz (solid triangles) are contrasted with the Weiner filter correlation functions for both cases, each for $\nu_{\Delta T} = 0, 1$. All but the unfiltered 90a+b GHz one agree.

| a+b ch | 53 | 53(400) | 53(200) | 53,+q | 53,+q(200) | 90 | 90,+q | 31 | firs |
|---|---|---|---|---|---|---|---|---|---|
| $10^5 \langle \mathcal{C}_\ell \rangle_{dmr}^{1/2}$ | 1.07±0.12 | 1.13±0.15 | 1.16±0.16 | 1.03±0.10 | 1.08±0.12 | 1.16±0.15 | 1.13±0.13 | 0.96±0.27 | 1.23±0.27 |
| $\nu_{\Delta T}+1$ | $1.4^{+0.4;0.7}_{-0.4;0.9}$ | $1.1^{+0.6;1.0}_{-0.4;0.7}$ | $1.0^{+0.5;0.9}_{-0.5;0.8}$ | $1.7^{+0.3;0.6}_{-0.4;0.8}$ | $1.3^{+0.5;0.9}_{-0.5;0.9}$ | $1.6^{+0.5;0.9}_{-0.5;1.0}$ | $1.9^{+0.4;0.8}_{-0.4;0.9}$ | $1.8^{+0.8;1.1}_{-1.0;1.6}$ | $1.6^{+0.7;1.2}_{-0.8;1.4}$ |
| channel | 53a | 53b | 53a−b | 90a | 90b | 90a−b | 31a | 31b | firs |
| ($\nu_{\Delta T}=0$) | 1.26±.12 | 1.06±.15 | $0.30^{+.22}_{-.30}$ | 0.82±.28 | 1.24±.21 | $0^{+.29}_{-0}$ | 0.82±.39 | 0.84±.45 | 1.15±.28 |
| ($\nu_{\Delta T}=1$) | 1.13±.11 | 0.97±.12 | $0.36^{+.16}_{-.25}$ | 0.91±.26 | 1.22±.17 | $0^{+.33}_{-0}$ | 0.86±.39 | 0.91±.47 | 1.28±.25 |

The Weiner filter depends upon the $\langle \mathcal{C}_\ell \rangle_{dmr}$ and $\nu_{\Delta T}$ we choose: I use the maximum likelihood value for $\langle \mathcal{C}_\ell \rangle_{dmr}$; and both maps and correlation functions are relatively insensitive to $\nu_{\Delta T}$ for large scale features. When the noise is large, as it is for these maps, it is the higher $\ell$ power that is preferentially removed by optimal filtering. Thus the theoretical fluctuation $\Delta s_k$ would have to be added to the Weiner-filtered maps of Fig.2 to give a realistic picture of the sky given the data and the theory. That is the maps are too smooth, 31 GHz the most, 53 GHz the least. The optimally-filtered correlation functions for *both* 53 and 90 GHz converge to the raw 53 GHz correlation function, and are $\nu_{\Delta T}$-independent (but this does not address how statistically significant such a $C(\theta)$ is for the given $\nu_{\Delta T}$). Not shown are the $C(\theta)$ for the optimally filtered *dmr* a−b difference maps, but these are nicely zero. Also not shown, but very gratifying, is that the main features are evident in both *A* and *B* channel maps, including in the noisy 90a and 31a,b.

Although correlation functions are quadratic combinations of pixel amplitudes and thus more complicated than the linear (Gaussian-distributed) statistics used here and in [14], they are effective filters of small-angle high-frequency systematic effects: a pixel-error enhancement ($r$) only contributes to the zero angle bin of $C(\theta)$, and the low $S/N$-modes at high $k$ have a range below about $\theta_{fwhm}$, the beamsize [6]. Thus $\nu_{\Delta T}$ determined with $C(\theta)$ should be a reasonable estimate if bins within $\theta_{fwhm}$ are not included (although this increases the error bars).

This letter shows that the overall anisotropy power COBE has discovered exists in all channels and is very robust. What drives the index $\nu_{\Delta T}$ high is more indicative of leftover residuals than suggestive of high-tilt primary anisotropies which are problematic for inflation-based models. Although more computer-intensive statistical exploration is still needed to determine the best $\nu_{\Delta T}$, especially when the four full years of *dmr* data are released, it seems unlikely the errors will ever be small enough that a strong argument against inflation based solely on *dmr* can be made. This research was supported by NSERC and the Canadian Institute for Advanced Research, and owes much to the nice job the *dmr* team did with the two-year data set.